\documentclass[a4paper]{article}
\usepackage[T1]{fontenc} %
\usepackage{RR}
\usepackage{hyperref}
\usepackage{color}
\usepackage{cite}
\usepackage{graphicx}
\usepackage{algorithm}
\usepackage{algorithmic}
\usepackage{ntheorem}

\usepackage[cmex10]{amsmath}

\newcommand{\power}{P_T}
\newcommand{\T}{\mathcal{T}}
\newcommand{\M}{\mathcal{M}}
\newcommand{\X}{\mathcal{X}}
\newcommand{\V}{\mathcal{V}}
\newcommand{\Orig}{\mathcal{O}}
\newcommand{\D}{\mathcal{D}}
\newcommand{\E}{\mathcal{E}}
\newcommand{\R}{\mathcal{R}}
\newcommand{\Nout}[2]{ \overrightarrow{ \mathcal{N}_{#1}^{#2} } }
\newcommand{\Nin}[2]{ \overleftarrow{ \mathcal{N}_{#1}^{#2} } }
\newcommand{\Fout}[3]{\overrightarrow{ F_{#1}^{#2}}(#3)} 
\newcommand{\Mmatrix}{M}
\newcommand{\Qmatrix}{Q}
\newcommand{\Rmatrix}{D}
\newcommand{\Nmatrix}{M_F}
\newcommand{\Smatrix}{S_M}
\newcommand{\Msrc}{\Mmatrix_S}
\newcommand{\Qsrc}{\Qmatrix_S}
\newcommand{\Rsrc}{\Rmatrix_S}
\newcommand{\Qe}[4]{\Qmatrix_{#1#2}^{#3#4}} 
\newcommand{\Qd}[4]{\Rmatrix_{#1#2}^{#3#4}}

\RRdate{September 2010}

\RRauthor{
Katia Jaffr\`es-Runser
  \thanks[sfn1]{\'Equipe SWING} 
  \thanks[sfn2]{Stevens Institute of Technology, Hoboken, NJ, USA}%
  \and
Cristina Comaniciu
\thanksref{sfn2}
    \and
Jean-Marie Gorce
\thanksref{sfn1}
}
\authorhead{Jaffr\`es-Runser, Gorce \& Comaniciu}
\RRtitle{De la prise en compte des interf\'erences et de la nature broadcast du canal radio pour l'\'evaluation de performance des r\'eseaux sans-fils}
\RRetitle{On the performance evaluation of wireless networks with broadcast and interference-limited channels}
\titlehead{On the performance evaluation of wireless networks}
\RRnote{This paper has been submitted to IEEE INFOCOM 2011}
\RRresume{
Ce rapport de recherche pr\'esente un mod\`ele multiobjectif d'\'evaluation des performances d'un r\'eseau ad hoc sans-fils. Les crit\`eres tels que la capacit\'e, la robustesse, l'\'energie et le d\'elais de transmission sont optimis\'es conjointement.
Gr\^ace \`a ce mod\`ele il est possible de d\'eterminer une borne en performance Pareto-optimale et les diff\'erents param\`etres du r\'eseau qui permettent d'obtenir de telles bornes. 
L'originalit\'e de cette approche r\'eside dans le fait qu'elle mod\`ele finement la nature broadcast intrins\`eque du canal radio et prend en compte de fa\,con pr\'ecise la distribution des interf\'erences dans le r\'eseau. 
Il est ici possible d'optimiser les d\'ecisions de routage et d'ordonnancement des paquets dans un r\'eseau quand plusieurs flots sont transmis dans le r\'eseau.
La complexit\'e de calcul de l'ensemble des solutions Pareto-optimale n'augmente |pas avec le nombre de flots pr\'esents dans le r\'eseau. 
La contribution majeure de ces travaux est la pr\'esentation d'un nouvelle formulation analytique des mesures de performance. 
Cette formulation se base sur une repr\'esentation matricielle des contraintes impos\'ees par la nature broadcast du canal et la distribution des interf\'erences dans le r\'eseau. 
Du fait de la similarit\'e de cette matrice avec une matrice Markovienne, il est possible d'exploiter des r\'esultats classiques de la th\'eorie des cha\^ines de Markov pour d\'eduire des m\'etriques de capacit\'e, robustesse, \'energie et d\'elais, le tout pour un r\'egime permanent.  
}
\RRabstract{
In this report we propose a MultiObjective (MO) performance evaluation framework for wireless ad hoc networks where criteria such as capacity, robustness, energy and delay are optimized concurrently.
Within such a framework, we can determine both the Pareto-optimal performance bounds and the networking parameters that provide these bounds. 
The originality of this approach is that it accounts for the inherent broadcast properties of the transmission and finely models the interference distribution.
In the proposed model, the network performance can be optimized when several flows (source-destination transmissions) exist. One benefit of our approach is that the complexity does not grow with the number of flows.
The other major contribution of this paper is the new analytical formulation of the performance metrics. It relies on a matrix representation of the constraints imposed by the interference-limited and broadcast wireless channel. Because of the similarity of this matrix with a Markovian transition matrix, we can exploit classical results from Markov chains theory to derive steady state performance metrics relative to capacity, robustness, energy and delay.    
Another very interesting feature of these new metrics is that the Pareto-optimal solutions related to them provide a tight bound on capacity, robustness, energy and delay.  
}

\RRmotcle{r\'eseaux ad hoc sans-fil, optimisation multi-objectifs, \'evaluation de performances}
\RRkeyword{wireless ad hoc networks, multiobjective optimization, performance evaluation}
\RRprojets{SWING}
\RRdomaine{1} 
\RRtheme{Mod\`ele pour l \'evaluation des performance d'un r\'eseaux ad hoc sans-fils}
\URRhoneAlpes 
\RCGrenoble 

\begin{document}
\RRNo{7379}
\makeRR   

\section{Introduction}\label{sec:introduction}
Wireless ad hoc and sensor networks are many times operating in difficult environments and require several performance criteria to be satisfied, related to timely, reliable, and secure information transfer.Routing and resource allocation protocols are key elements  for ensuring the information transfer across the network.
To better understand the capabilities of such protocols for a given network topology, it would be very helpful to know the bounds that can be achieved with respect to multiple performance criteria. These bounds illustrate the interdependence between multiple performance metrics and can capture the tradeoffs between the various operating points for the network. As a consequence, defining a unified multi-objective design framework capable of capturing the above mentioned tradeoffs for various possible operating points for the network becomes of premier importance.

However, globally optimizing capacity for two-dimensional networks where routing and resource allocation (i.e. frequency, time or power assignment) are performed concurrently is a very hard problem that has triggered a comprehensive research effort under various conditions since the seminal work of Gupta and Kumar \cite{Gupta2000}. Several works provided ways for increasing the asymptotic capacity bound of $O(n~log(n))$ \cite{Gupta2000,Wang2008} by for instance accounting for mobility \cite{Grossglauser2002}.
In these works two important assumptions are made: unicast communications and threshold-based interference models.
In a unicast communication, packets are sent to a specific receiver on a multi-hop chain. In a threshold-based model, nodes interfere with other nodes within a fixed range while beyond that range, no node is interfered. 
These results have been extended in \cite{Mhatre2009} by accounting for realistic additive interference or in \cite{Comaniciu2006} by considering multi-user detection techniques. 
Toumpis et al. \cite{toumpis2003} have also proposed an interesting model to derive the capacity region by properly scheduling transmissions and hence accounting for a temporal multiplexing directly in a $2$ dimensional network model. While the model is appealing, it is not scalable with respect to the network size and the number of flows transmitted.

The broadcast nature of the wireless channel is mostly assumed as the main source of interference in these works and hence negatively impacts the network performance. However, this broadcast property can serve the capacity if the receiving nodes are able to cooperate and coordinate to transmit the data flow \cite{Dana2006}.  It has been shown to be beneficial in the context of opportunistic routing \cite{Jacquet-JSAC2009} as well.

In this paper we consider the broadcast nature of the channel as a possible way to increase the capacity, and we propose realistic interference models. This work does not aim to assess the performance of a specific routing protocol, but more generally to extract the boundary of the feasible performance region of the network. Further, we consider that capacity is not the only interesting objective. More specifically, it is the trade-off between different performance criteria that should be considered to properly select a transmission strategy in the network. 
For instance, increasing the number of parallel paths improves capacity but at the expense on an increased energy due to the multiplication of packets traveling in the network. Thus, various criteria related to transmission delay \cite{Brand2008r}, energy consumption \cite{Vassileva2007} or fairness \cite{Eryilmaz2006} should be considered in addition to the main design goal of reliable information transmission. 
As a consequence, the assessment of networking protocols usually relies on various criteria which may be evaluated analytically or through network simulations.
In this work, we concentrate on three criteria: capacity, energy and latency. As we will see in Section \ref{sec:steadystate}, robustness is related to capacity with a redundancy factor. 

We consider a network as a set of nodes comprised of relays, sources and receivers. In this work, we do not look for the optimal single path between a source and a destination, but for a set of relays deciding to forward packets or not. This approach is more closely related to a diffusion mechanisms than to a strict single path routing protocol. 
However, our more general formulation encompasses single path routing solutions. 
This concept has been inspired by the pioneering works on optic and electrostatic inspired routing \cite{Jacquet2004-MobiHoc,Toumpis2008-ComputNet,Altman2008-AdhocNow}. 
However, our work is not directly related to a physics inspired phenomenon. 
We provide here a more practical framework that relies on a time and spatial multiplexed approach: we model the decision of each relay to transmit, receive or just sleep with a given probability associated to a time slot. These forwarding probabilities define a steady state of the network.
Then, we propose an analytical framework to obtain an estimate of overall capacity, latency and energy depending on the forwarding probabilities associated to each node and time slot. 
 Since interference is modeled for a network steady state, the results obtained scale with respect to the number of flows being transmitted in the network. 

The paper is organized as follows. Section \ref{sec:preliminaries} gives some preliminary notations. The network model is detailed in Section \ref{sec:networkmodel} and Section \ref{sec:problemstatement} describes the MO optimization problem. The steady-state performance evaluation tools are described in Section \ref{sec:steadystate} and Section \ref{sec:conclusion} concludes the paper. 

\section{Preliminaries}\label{sec:preliminaries}

\subsection{Notations}
Throughout this paper, upper case letters (e.g., $X$, $Y$, $Z$) usually denote vectors or matrices. 
Lower case letters (e.g., $x$, $y$, $z$) denote probability values. 
The transpose of a vector or a matrix is shown by $F^{\dagger}$. 
Subscripts denote the indeces of nodes of the network. For instance, $F_i$ could denote a vector relative to node $i$ and $p_{ij}$ a probability related to both nodes $i$ and $j$.
Superscripts are introduced to specify the index of a time slot. For instance, $x^{uv}$ could denote a probability value relative to time slot $u$ and $v$. 
Sets are denoted by calligraphic alphabets (e.g., $\mathcal{A}$, $\mathcal{B}$, $\mathcal{C}$) and the cardinality of set $\X$ is denoted by $|\X|$. The complement of a set $\mathcal{A}$ is shown by $\mathcal{A}^{c}$.

A flow of symbols coming \emph{out} of a node $i$ on time slot $u$ is denoted by $\overrightarrow{f_i^u}$ and a flow of symbols coming \emph{into} node $i$ from node $k$ in time slot $v$ is denoted by  $\overleftarrow{f_{ki}^v}$.  
Table~\ref{tab:notations} summarizes our notations.

\begin{table} 
\caption{Some important notations in this paper} 
\centering
\label{tab:notations}
\begin{tabular}{| c | l |}
\hline
$\V$			& Set of nodes \\
$\E$		& Set of edges \\
$\Orig$		& Set of source nodes \\
$\D$		& Set of destination nodes\\
$\R$		& Set of possible relay nodes\\
$N$				& Number of possible relay nodes\\
$\T$				& Set of time slots composing one frame\\
$\Nout{i}{u}$ 		& Set of incoming edges at node $i$  on time slot $u$\\
$\Nin{i}{u}$ 		& Set of outcoming edges at node $i$ on time slot $u$\\  
$\tau^u_i$		& \emph{Transmission rate} for node $i$ in time slot $u$ \\
$\Delta$			& Discrete set of possible values of  $\tau^u_i$\\			
$p_{ij}^u$ 		& \emph{Channel probability} on edge $(i,j)$ for slot $u$ \\ 
$x_{ij}^{uv}$		& \emph{Forwarding probability} for a packet sent by node $i$ in \\
				&  time slot $u$ to be forwarded by node $j$ in time slot $v$\\
$\M^u$			& Set of active transmission in time slot $u$\\
$\M$				& Set of all actives transmission \\
$f_R, f_D, f_E$		& Redundancy, delay and energy criteria, resp. \\
\hline

\end{tabular}
\end{table}

\subsection{Definitions of graph theory}

In this part, we briefly review the concepts and definitions from graph theory considered in this paper \cite{graph}. 
A complete graph $\mathcal{G=(V,E)}$ has vertex set $\V$ and edge set $\mathcal{E \subset V \times V}$. Without loss of generality, let
$ \V = \{1, 2, \dots, |\V| \} $.

We assume that the graph is finite, i.e., $|\V| \leq \infty$. 
For each node $i \in \V$, $\Nout{i}{}$ and $\Nin{i}{}$ are the set of edges leaving from and the set of edges going into $v$, respectively. Formally 
\begin{eqnarray*}
\Nout{i}{} 	& = &\{ (i,j) | (i,j) \in \E\}  \\
\Nin{i}{} 	& = &\{ (j',i) | (j',i) \in \E\}.
\end{eqnarray*}

A complete graph $\mathcal{K_{|\V|}}$ of $|\V|$ vertices is a graph having the maximum number of edges.


\section{Network model}\label{sec:networkmodel}

\begin{figure}
\begin{center}
  \scalebox{0.45}{
  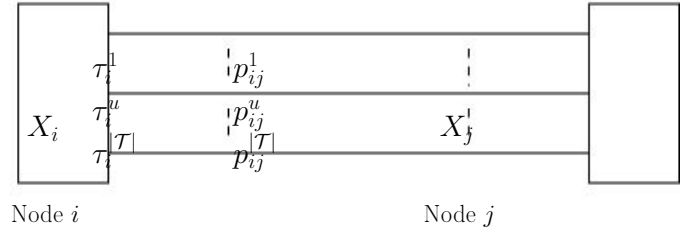}
  \caption{Relay network model: transmission rate and channel probabilities}
\label{fig:networkmodel}
\end{center}
\end{figure}

We model the wireless ad hoc network by a finite complete graph $\mathcal{K_{|\V|}}$ for two reasons: modeling the \emph{broadcast nature} of the wireless channel and finely accounting for \emph{interference}.
Transmission in the network is time multiplexed and a frame of $|\T|$ time slots is repeated in each epoch $s$. 
Each edge $(i,j) \in \E$ for a particular time slot $u$ represents an interference-limited channel which is modeled by the probability of a symbol or a packet to be correctly transmitted. This probability is referred to as the \emph{channel probability} in the following. It models interference as an additive noise and is computed considering the distribution of the bit error rates (BER) or the packet error rates (PER) as shown hereafter. As a consequence, there are  $|\T|$ orthogonal interference-limited channels for each edge $(i,j) \in \E$ as illustrated on Fig.~\ref{fig:networkmodel}.
Each channel is assumed to be in a hald-duplex mode, i.e. a node cannot transmit and receive a packet at the same time.

A set of sources $\Orig$ and destinations $\D$ is defined. A flow of packets is transmitted between any source-destination combination $(\{O_i, i \in \left[1,..,|\Orig| \right]\},\{D_j, j \in \left[1,..,|\Orig| \right]\})$. 
We make the assumption that source and destination nodes do not relay the information. As a consequence, the network we are modeling is composed of a set of relay nodes $\R = \V-\Orig-\D$. In the following, we consider that the number of relays in the network is $N = |\R|$. We also assume that a relay can not differentiate packets. As a consequence, all packets are treated as being unique by a relay.

A \emph{transmission} is defined as the couple $(i,u) \in \V \times \T$ and represents the fact that node $i$ is transmitting in a time slot $u$.

\subsection{Transmission rate}
We consider that a node $j$ transmits a flow of symbols in time slot $v$. The symbols are the realization of a random variable $\overrightarrow{X_j^v}$ chosen from an alphabet $\X$ ($\X=\{0,1\}$ for instance). With this definition, we consider that a node $j$ transmits the same symbol in a time slot $v$ on all its outgoing edges $\Nout{j}{v}$. This constraint incorporates the broadcast property of wireless communications. Time multiplexing is here accounted for in our model since a node can transmit different symbols on each available time slot. 
 
A flow of symbols  coming into node $j$ from node $i$ on time slot $u$ is modeled as a random variable $\overleftarrow{Y_{ij}^u}$. The symbols are transmitted by node $i$ on the edge $(i,j) \in \E$ in time slot $u$ and consequently, experience the probability $p_{ij}^u$ of being received successfully in $j$. Depending on the interference temperature of the network, symbols are received successfully or not at $j$. 

Having this, let $\overleftarrow{Y_j} = \{ \overleftarrow{Y_{ij}}, (i,j) \in \Nin{j}{}\}$ be the random variable giving all the symbols that can be received on the incoming channels of node $j$ for all time slots where $\overleftarrow{Y_{ij}}=\{\overleftarrow{Y_{ij}^u}, (i,j) \in \Nin{j}{u}\}$. 
Let $\overrightarrow{X_j}=\{\overrightarrow{X_j^v}, j \in \Nout{j}{v}\}$ be the random variable giving all the symbols that can be transmitted on the outgoing channels of node $j$ for all the time slots. 
The relation between the \emph{outgoing} random variables $\overrightarrow{X_j}$ and the \emph{incoming} random variables  $\overleftarrow{Y_j}$ defines a coding scheme for the network.
 
A node has one main decision to take upon receiving a symbol or a packet: whether it should process it or not. If it decides to process it, the next steps are to

$\bullet$ Decide on the coding for the symbol or packet to be sent. 

$\bullet$ Decide on which channel to transmit it. 

These decisions strongly influence the quality of the transmission and aim at mitigating the transmission errors due to fading and interference. 
Depending on these decisions, the transmission rate of a node $i$ on a channel $u$ varies. For instance, if a node decides to drop one packet out of two received on a same time slot $u$, the transmission rate on channel $u$ would become half the rate at which it received packets on the same time slot. 
For instance, it is possible that a node transmits one packet every two received packets on time slot $u$ because it is applying a coding scheme for which two packets are combined into a single transmitted packet. In both cases, the node is adapting the transmission rate on each channel to combat errors.   

Based on this statement, we characterize the behavior of a node $i$ by the rate at which it is transmitting in each time slot $u$. The \emph{transmission rate} of node $i$ in time slot $u$ is denoted by $\tau_i^u$. Having this, a vector of transmission rates for each time slot can be defined
\[ \tau_i = \left[ 
	\begin{array}{ccc}
	\tau_i^1&
	 \dots1& 
	 \tau_i^{|\T|} \\
	 \end{array}
	 \right]
,~ \tau_i^u \in [0,1]
\]  

The transmission rate can be interpreted as the percentage of time a node is transmitting packets in a given time slot. For instance, if a node $i$ transmits at a rate of $\tau_i^u=0.5$ in time slot $u$, it transmits a packet every two time slots, if one time slot permits to transmit exactly one packet. 
We assume that when a node is not transmitting on a channel, which happens with probability $1-\tau_i^u$, it is listening to packets. This is to comply with the half duplex assumption. In other words, the \emph{listening rate} is equal to $1-\tau_i^u$. The model could be extended to also account for the time a node may be sleeping, but this option has been disregarded so far because it would increase the dimension of our search space. However, it would be worth investigating such an option when minimal energy consumption prevails as it is the case for wireless sensor networks.
The proposed network model is consequently defined in a steady state mode, where we know which nodes are transmitting or not packets, continuously.

The average rate at which all the symbols are coming into node $j$ is given by 
\[ 
\overleftarrow{r_j} = \sum_{u \in \T} \sum_{(i,j) \in \Nout{j}{u}} \tau_i^u p_{ij}^u
\]
for $p_{ij}^u$ the probability to receive a symbol correctly. 
The rate at which symbols are being transmitted by node $j$ is given by: 
\[ 
\overrightarrow{r_j} = \sum_{v \in \T} \tau_j^v
\]

Let \[\tau = \left[ 
 	\begin{array}{c c c}
		\tau_1^\dagger &
		\cdots &
		\tau_N^\dagger \\
 	\end{array}
\right]^{\dagger}
\] be the matrix of the transmission rates assigned to all the nodes of the network for all time slots. A particular instance of $\tau$ belongs to the set of possible transmission rate matrices shown by $\Gamma$.
$\tau$ is feasible if Properties 1 and 2 hold for each node: 

{\sc Property 1:} {\it Flow conservation}.  The rate of all outgoing flows is lower or equal to the rate of all incoming flows, i.e.
\begin{equation}
  \overrightarrow{r_j} \leq \overleftarrow{r_j}  
\label{eq:ratesflowconservation}
\end{equation}  

{\sc Property 2:} {\it Half duplex.} A node $j$ is able to receive a message on a time slot $u$ the proportion of time it is not transmitting on that same time slot. 
As a consequence, the proportion of packets being actually received on time slot $u$ is given by
\begin{equation}
\overleftarrow{r_j^u}(1-\tau_j^u) +  \tau_j^u \leq 1, ~~\forall u \in \T
\label{eq:halfduplex}
\end{equation}  
\noindent where $\overleftarrow{r_j^u} = \sum_{(i,j) \in \Nout{j}{u}} \tau_i^u p_{ij}^u$ stands for the incoming rate in time slot $u$. The conditions of \eqref{eq:halfduplex} are always true for the definition of the transmission and listening rates. However, this constraint becomes more meaningful if a sleeping rate variable is defined for a node. 
  
Now that we have defined the transmission rate matrix, we can define the set of active transmissions $\M$ corresponding to $\tau$. A transmission $(i,u) \in \V \times \T$ is said to be active if $\tau_i^u > 0$. As a consequence, the set $\M$ is given by:
\[
\M = \{(i,u) \in \V \times \T~ |~ \tau_i^u > 0\}
\]
Similarly, the set $\M^u$ refers to the set of transmission being active in a same time slot $u$. We have $\M^u = \{(i,v) \in \V \times \T | \tau_i^u > 0, v=u \}$

\subsection{Channel probability}
Knowing $\tau$, it is straightforward to derive the interference temperature of the network. We model interference as an additive noise computed at the end of each link $(i,j) \in \E$ of the network. 
Let $p_{ij}^u$ be the probability for a symbol transmitted by node $i$ to arrive \emph{successfully} at node $j$ in time slot $u$. We denote $p_{ij}^u$ as the channel probability.
It is a function of the statistical distribution of the Signal to Noise and Interference Ratio (SINR) at the location of the destination node $j$. 
This quantity can either be defined using a Packet Error Rate (PER) or a Bit Error Rate (BER) depending if transmissions on the network are packetized or not.  
In the following we consider that transmissions are packetized.
Before giving the exact expression of the channel probability, a few preliminary definitions and notations are given hereafter:

\paragraph*{Pathloss attenuation factor and transmission power}
$a_{ij}$ reflects the attenuation due to propagation effects between nodes $i$ and $j$. In our simulations, the simple isotropic propagation model is considered.
We consider that all the nodes use the same transmission power denoted as $\power$.

\paragraph*{Interference}
Since we consider time-multiplexed channels, interference only occurs between transmissions using the same time slot.
Let  $I_{ij}^u$ be the power of the interference on the link $(i,j) \in \E$ on time slot $u$ and computed at node $j$. 
If we denote by $\mathcal{I}_{ij}^u$ a set of nodes of the network interfering at node $j$ (not including $i$) in time slot $u$, interference power is defined by:
\begin{equation}
\label{eq:interf}
I_{ij}^u = \sum_{k \in \mathcal{I}_{ij}^u} \power \cdot a_{Id(k)j}~~{\rm for}~\mathrm{Id}(k) \neq i
\end{equation}
\noindent where $Id(k)$ gives the number of the node interfering at $j$.
 
\paragraph*{SINR}
The SINR between any two nodes $i$ and $j$ in resource $u$ is given by the following equation:
\begin{equation}
\gamma_{ij}^u=\displaystyle{\frac{\power \cdot a_{ij}}{N_0+I_{ij}^u}}
\label{eq:sinr}
\end{equation}
where $I_{ij}^u$ is the interference power on the link and $N_0$ the noise power density.

\paragraph*{Packet error rate (PER)}
For a specific value of SINR $\gamma$, the packet error rate $PER$ can be computed according to:
\begin{equation}
PER(\gamma) = 1-\left[1-BER(\gamma)\right]^{N_b} 
\label{eq:PER}
\end{equation}
\noindent where $N_b$ is the number of bits of a data packet and $BER(\gamma)$ is the bit error rate for the specified SINR per bit $\gamma$ which depends on the physical layer technology and the statistics of the channel.
Results are given for an AWGN channel and a BPSK modulation without coding where $BER(\gamma)= Q\left( \sqrt{2\gamma}\right)= 0.5 *{\rm erfc}(\sqrt{\gamma})$.

\paragraph*{Interfering sets}
A node $i$ is said to be active in the network if $\tau_i^u>0$. We recall that $\M^u$ designates the set of active transmissions in time slot $u$. 
We give now a more precise definition of an interfering set. An interfering set $\mathcal{I}_{ij}^u(K)$ on a link $(i,j)$ in time slot $u$ is any subset of the set $\M^u-\{i\}$ made of $K$ elements, $K = \{1, .., |\M^u-\{i\}|\}$.
If $M=|\M^u-\{i\}|$, let $\mathcal{L}_{ij}^u$ refer to the set of all possible interfering sets of cardinality $|\mathcal{L}_{ij}^u|=\sum_{K=1}^{M}\left(^{M}_{~K}\right)+1$.

\vspace{\baselineskip}

Equation \eqref{eq:linkProbaPER} details the derivation of the channel probability $p_{ij}^u$ as the average of the PER experienced for all possible interfering sets $l \in \mathcal{L}_{ij}^u$ on time slot $u$ referred to as $PER_l$:
\begin{equation}
p_{ij}^u = \sum_{l \in \mathcal{L}_j^u} \left[ 1-PER_l\right].\mathbf{P}_l 
\label{eq:linkProbaPER}
\end{equation}
\noindent where $PER_l = PER(\gamma_l), l \in \mathcal{L}_{ij}^u$, where $\gamma_l$ is the SINR experienced on the link $(i,j)$ in time slot $u$ if the nodes of interfering set $l$ are active.  

$\mathbf{P}_l$ is the probability for the interfering set $l$ to be active and create interference on the link $(i,j)$ on time slot $u$. More specifically, it is the probability that the nodes of the interfering set $l$ are transmitting concurrently and the others are not as specified in the following:
\begin{equation}
\mathbf{P}_l = \prod_{k=1}^K \tau_k^u~\cdot~\prod_{m=1}^{M-K-1}(1-\tau_m^u)
\label{eq:probaPER}
\end{equation}
In \eqref{eq:probaPER}, $\prod_{k=1}^K \tau_k^u$ gives the probability that the $K$ active nodes of the interfering set $l$ are transmitting and $\prod_{m=1}^{M-K-1}(1-\tau_m^u)$ the probability that the $M-K-1$ other active nodes are not.

Similarly to the vector of transmission rates, we define the vector of channel probabilities for a link $(i,j) \in \E$:
\[ P_{ij} = \left[
	\begin{array}{c c c}
		p_{ij}^1 & \cdots & p_{ij}^{|\T|} \\
	\end{array}
\right]
, p_{ij}^u \in \left[0,1\right] \]

Let $P_j$ be the matrix giving all incoming channel probabilities at node $j$
\[
P_j = \left[ 
 	\begin{array}{c c c}
		P_{1j}^\dagger &
		\cdots &
		P_{Nj}^\dagger \\
 	\end{array}
\right]^{\dagger}
\]
and $P$ the matrix of all channel probabilities in the network 
\[
P = \left[ 
 	\begin{array}{c c c}
		P_{1} &
		\cdots &
		P_{N} \\
 	\end{array}
\right]
\]

\subsection{Forwarding and scheduling}

Now, lets introduce the forwarding and scheduling decisions of the nodes relative to the transmission of a packet. This decision is represented by the probability $x_{ij}^{uv}$ of a node $j$ to transmit on time slot $v$ a packet coming from node $i$ on time slot $u$. We will refer in the rest of the text to the forwarding probability  $x_{ij}^{uv}$. 
For each node of the network, we can define a $(N.|\T|)$-by-$|\T|$ matrix giving all the forwarding probabilities relative to any node $j$ of the network as follows. This forwarding matrix $X_j$ is given by:

\[
X_j = \left[ 
 	\begin{array}{c c c}
		X_{1j} &
		\cdots &
		X_{Nj} \\
 	\end{array}
\right]^{\dagger}
\]
where each matrix  $X_{ij}$ provides the scheduling probabilities of a flow of packets coming from node $i$ on its output times slots, depending on the time slot the packets are received on.
\[ 
 X_{ij} = \left[ 
 	\begin{array}{c c c}
		x_{ij}^{11} 	& \cdots & x_{ij}^{1|\R|}\\
		\vdots 	 	& 		&	\vdots	\\
		x_{ij}^{|\R|1} 	& \cdots & x_{ij}^{|\R||\R|}\\
 	\end{array}
\right] 
\]

The matrix of forwarding probabilities is related to the matrix of transmission rates $\tau$ and the matrix of channel probabilities $P$ with the following set of $|\M|$ equations
\begin{equation}
\sum_{(i,j) \in \Nout{j}{u}} \sum_{u \in \T}  \tau_i^u p_{ij}^u (1-\tau_j^v) x_{ij}^{uv} =  \tau_j^v, ~~~\forall (j,v)\in \M
\label{eq:fwfprobaflowconservation}
\end{equation}
\noindent where $\tau_i^u p_{ij}^u$ is the probability that a packet sent by $i$ on time slot $u$ arrives and $ (1-\tau_j^v)$ is the probability that node $j$ is listening on channel $v$. These equations introduce strict constraints on the choices of the forwarding probabilities. 

The forwarding probabilities represent the decisions of the nodes to either $(i)$ retransmit all the packets or symbols received or $(ii)$ reduce the output rate by dropping or re-encoding them together. 
From now on, we will refer to the set of all forwarding probabilities of the complete network using a matrix $X$ of size $N.|\T|$-by-$N.|\T|$ defined by:

\begin{equation}
X = \left[ X_1 \dots X_N \right],~ X \in \X 
\label{eq:xmatrix}
\end{equation}
\noindent where $\X$ is the set of all possible matrix instances.


\section{MO optimization problem}\label{sec:problemstatement}

\begin{figure}
\begin{center}
\scalebox{0.5}{
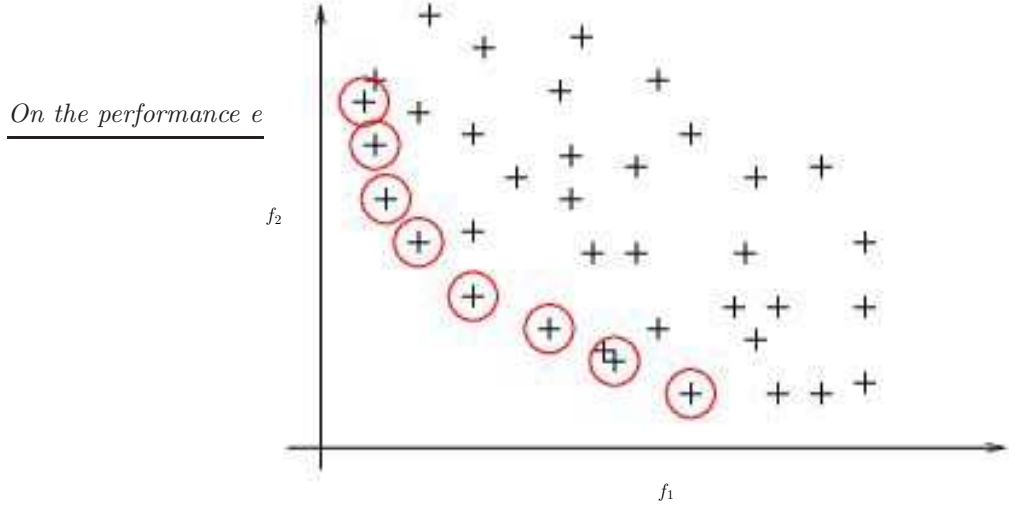}
\caption{Non-dominated solutions are highlighted with red circles for a 2-function minimization problem.}
\label{fig:dominance}
\end{center}
\end{figure}

The scope of this section is now to take advantage of the previously described network model to derive a framework capable of extracting the set of Pareto-optimal transmission strategies with respect to various performance criteria (e.g. capacity, delay, energy$\dots$ ).
A Pareto-optimal set is composed of all the non-dominated solutions of the MO problem with respect to the performance metrics considered. The definition of dominance is:

{\sc Definition 1:} A solution $A$ dominates a solution $B$ for a $n-$objective MO problem if $A$ is at least as good as $B$ for all the objectives and $A$ is strictly better than $B$ for at least one objective. Mathematically, we have for a minimization problem:
\begin{equation}
\label{eq:dominance}
\begin{split}
A \succ B \equiv  \\ \forall i \in [1,n] : f_i(A)\leq f_i(B), \exists j \in [1,n] : f_j(A) < f_j(B)
\end{split}
\end{equation}
\noindent Notation $A \succ B$ means that $A$ strictly dominates $B$.
In the space of the evaluation functions, for the case of a minimization of all the optimization objectives, the set of Pareto-optimal solutions for an example 2-objective problem is illustrated in Fig.~\ref{fig:dominance}.

\subsection{Solution definition}
A solution of our MO optimization problem is given by the set of all the forwarding probabilities represented by matrix $X \in \X$ defined in \eqref{eq:xmatrix}. As a consequence, the set $\X$ becomes the problem search space. 
Let  $f_C, f_R, f_D$ and $f_E$ be the performance criteria relative to capacity, robustness, delay and energy, respectively. Capacity and robustness are maximized, while energy and delay are minimized. The derivation of these metrics is provided in the next section. 
Our goal is to solve the following multiobjective optimization problem by finding the set of Pareto-optimal solutions $\X_{opt}$:
\begin{equation}
\X_{opt} = \{A \in \X~| ~\forall A \in \X_{opt}, \forall B \in \X_{opt}^{c},  A \succ B \}
\label{eq:MOproblem}
\end{equation}
\noindent where $  \X_{opt} \cup \X_{opt}^{c} = \X$. The dominance relation is defined in \eqref{eq:dominance} and adapted for criteria $f_C, f_R, f_D$ and $f_E$.

The cardinality of the search space $\X$ as defined grows exponentially with $N$ and $|\T|$. However, considering the complete search space may be meaningless depending on the network connectivity.  
As a consequence, it is reasonable to define the search space for a maximum number of relays $N_{max}<N$.     
That is why we define $\X_{N_{max}}$ as the subset of $\X$ made of at most $N_{max}$ active relays and refer in that case to the $N_{max}$-relay problem. In this case, we are looking for the Pareto-optimal combinations $X$ of $N_{max}$ relays in the set $\R$.  

There is a whole set of constraints for a solution $X \in  \X_{N_{max}}$ to be valid related to Properties 1 and 2. As a consequence, the solutions of   $\X_{N_{max}}$ that do not follow these properties are dropped before being evaluated by the search algorithm since they are considered as not feasible. We recall that Property 1 ensures that the flow conservation constraint at each node holds and that Property 2 ensures that the half duplex constraint is met.  

\subsection{Change of variable}

The derivation of the transition rate matrix $\tau \in \Gamma$ knowing a solution $X \in \X$ is intractable. In a nutshell, to compute the transmission rates of a node $j$, $X$ and the incoming transmission probabilities $P_j$  have to be known yet the elements of $P_{j}$ are a function of the transmission rates matrix, creating a circular dependency between the variables.

Nevertheless, being able to model exactly the interference temperature of the network is appealing and to keep the same formalism, we propose a reverse approach where a solution of the optimization problem is defined by the set $\tau$ of the transmission rates for all the nodes and time slots. From $\tau$, it is possible to derive the channel probabilities matrix $P$ according to \eqref{eq:probaPER} since the activity of all nodes on each time slot is known. 

Only instances of $\tau$ that meet the constraints relative to Property 1 and 2 are further considered as valid. 
Now that we have a valid $\tau$, we can derive all the forwarding matrices $X \in \X$ that verify the constraints of equation \eqref{eq:fwfprobaflowconservation}. 
There are $|\M|$ constraints, each one constraining the choice of the $x_{ij}^{uv}$ for all nodes and time slots of the network with respect to $\tau$.
Let $\X^{\tau}$ be the subset of $\X$ that verifies \eqref{eq:fwfprobaflowconservation} with respect to the transmission rate matrix $\tau$. 
Each solution $X \in \X^{\tau}$ can be evaluated according to $f_C, f_R, f_D$ and $f_E$. 

The MO optimization problem of \eqref{eq:MOproblem} stays unchanged, however, the way the search space is constructed has changed and permits to select in a first stage a subset of valid and interesting solutions $\X^{\tau}$. 
This feature is very interesting in terms of optimization. In fact, if it is possible to determine knowing $\tau$ if the set of corresponding $\X^{\tau}$ is promising or not, it is possible to apply pruning techniques that disregard non-promising $\tau$'s instances. 

\subsection{Problem complexity}
We recall that each transmission rate $\tau_i^u$ takes its values in the continuous closed set $\left[0,1\right]$. However, to be able to use common MO optimization heuristics, we formulate the problem as a combinatorial MO optimization problem. Therefore,  transmission rates take their values in a discretized set $\Delta = \{\delta_1, \dots, \delta_{|\Delta|}\}, \delta_i \in \left[0,1\right]$ of $|\Delta|$ values.

The set $\Gamma$ of possible $\tau$ solutions has a cardinality of $|\Gamma|=|\Delta|^{N.|\T|}$. The set of feasible solutions is reduced by Property 1 but since the relation between $\tau$ and $P$ is complex, it is difficult to derive analytically the exact number of feasible solutions. We can just say that $\Gamma$ is bounded by $O(|\Delta|^{N.|\T|})$.

The set $\X$ of possible forwarding probabilities matrices has a cardinality of $|\X|=|\Delta|^{2N(N+|\Orig|-1).|\T|}$. The subsets $\X^\tau$ of feasible $X$ variables is also reduced because of Property 1. Let $\X_{f}$ be the set of feasible $X$ solutions. 
$\X_{f}$ is the union of the $\X^\tau$ for all feasible $\tau$ instances. 
Similarly to $\Gamma$, it is difficult to evaluate the size of $\X_f$. Thus, $\X_f = O(|\Delta|^{N^2.|\T|})$.


\section{Steady state performance evaluation}\label{sec:steadystate}

This section provides a framework for the definition of performance criteria for a particular solution $X \in \X$ of our MO optimization problem. 
It is an important contribution because it permits a fast performance derivation for end-to-end criteria such as capacity, robustness, delay and energy. Moreover, this framework also permits to reduce the complexity of the problem by pruning solutions where the transmission rate matrix $\tau$ does not meet sufficient performance constraints. 
Our framework relies on the definition of a transition matrix which is composed of the probabilities for a flow of packets to be re-transmitted by the nodes of the network. Such a formulation is inspired by the theory of Markov chains. However, we want to stress that we do not model the network using a Markov chain. We define instead a transition matrix that has the properties needed to be able to re-use some results from the theory of Markov chains.
Table \ref{tab:notations2} gives notations specific to this section. 

\begin{table} 
\caption{Main notations in this section} 
\centering
\label{tab:notations2}
\begin{tabular}{| c | l |}
\hline
$\Fout{out}{}{s}$	& Flow vector at time epoch $s$ \\
$\Mmatrix$		& Transition matrix \\
$\Qmatrix$		& Relaying matrix \\
$\Rmatrix$		& Arrival matrix \\
$\Nmatrix$		& Fundamental matrix \\
$\Smatrix$		& Source matrix \\
\hline

\end{tabular}
\end{table}

\subsection{Flow vector}
Let $\Fout{i}{u}{s}$ represent the probability of node $i$ to transmit a packet in time slot $u$ at time epoch $s$. We consider that at the beginning of a time epoch, a frame of $|\T|$ time slots starts and all nodes having a packet to send can transmit it in their respective time slots. A new epoch starts when all nodes have finished receiving their packets and are ready to send new ones if needed. 
Let $\Fout{out}{}{s}$ be a vectorial representation of these probabilities for each node and each time slot at a given time epoch $s$. 
$\Fout{out}{}{s}$ is referred to as the \emph{flow vector} and is shown by:
\[
\Fout{out}{}{s} = 
\left[ 
 	\begin{array}{c c c  c}
		\Fout{1}{}{s} & 
		\cdots  & 
		\Fout{N}{}{s}, &
		\Fout{D}{}{s}\\
 	\end{array}
\right]
\]
\noindent where $\overrightarrow{F_i}(s) = \left[ \begin{array}{ccc} \Fout{i}{1}{s} & \cdots & \Fout{i}{|\T|}{s} \end{array} \right]$.

The vector $\Fout{D}{}{s}$ stands for the flows received at the destinations nodes of set $\D$. 
It is given by $\Fout{D}{}{s}=\left[ \begin{array}{ccc} \Fout{D_1}{}{s} & \cdots & \Fout{D_{|\D|}}{}{s} \end{array} \right]$ where $\Fout{D_i}{}{s}$ gives the flows received in each time slot by destination $D_i$ .

The flow vector is a $(N+|\D|).|\T|$ matrix when all the possible transmissions are accounted for. However, if a relay $i$ never transmits in a time slot $u$ (i.e. $\tau_i^u=0$), we have $\Fout{i}{u}{s}=0$.    
As a consequence, the size of the flow vector can be reduced to the size of the only transmissions that are active. Hence, the size of $\Fout{out}{}{s}$ can be reduced to a $|\M|+|\D|.|\T|$.

\subsection{Transition matrix}
The propagation of the packets and the forwarding decisions of the nodes can be represented by a transition matrix $M$.
Each element of the matrix $M$ gives the probability for a packet transmitted by one relay in one of the time slots to be transmitted by any other relay in the network in one of the time slots.   
When it is applied to the flow vector of time epoch $s$, the new flow vector of time epoch $(s+1)$ is obtained. Formally we have:
\[ \Fout{out}{}{s+1} = \Fout{out}{}{s}\cdot \Mmatrix
\]  

The transition matrix $\Mmatrix$ has the following structure:
\[\Mmatrix = \left[ 
	\begin{array}{c|c}
	 \Qmatrix   	& \Rmatrix \\ \hline
  	 0 	& I \\
	\end{array}
\right]\] 
\noindent For $l=N.|\T|$ and $m=|\D||\T|$, $\Qmatrix$ is a non-zero $l$-by-$l$ matrix, $\Rmatrix$ is a non-zero $l$-by-$m$ matrix, $I$ is a $m$-by-$m$ identity matrix and $0$ a  $m$-by-$l$ zero matrix. $\Qmatrix$ is referred to as the relaying matrix and $\Rmatrix$ as the arrival matrix. 

Transition matrix $\Mmatrix$ has a similar canonical structure than a finite absorbing Markov chain \cite{markov}. Here, we have $l$ transient states, i.e. relay nodes that are forwarding packets using active transmissions. The relaying matrix $\Qmatrix$ gives the probabilities for the packets sent at time epoch $s$ to be retransmitted at time epoch $s+1$ by the relays of the networks. 
We have $m$ absorbing states where the destination nodes of $\D$ receive packets on each time slot and keep them. The identity matrix represents the fact that packets received by a destination are never forwarded, but absorbed.  Matrix $\Rmatrix$ is composed of the probabilities from going from one transient state to one absorbing state, i.e. the probability for packets to be received at the destinations when being transmitted from any relay or source in the network. 

\emph{The relaying matrix $\Qmatrix$} is structured as follows:
\[
\Qmatrix = \left[
\begin{array}{cccc}
0			& \Qe{1}{2}{}{}	&	\cdots	& \Qe{1}{N}{}{} \\
\Qe{2}{1}{}{}	& 0			&	\cdots	& \Qe{2}{N}{}{} \\
\vdots		& 			&			& \vdots \\
\Qe{N}{1}{}{}	& \cdots		&	 \Qe{N-1}{N}{}{} 	& 0\\
\end{array}
\right]
\]
\noindent 0 is an $|\T|$-by-$|\T|$ zero matrix representing the fact that a node $i$ does not forward a packet coming from itself. The matrix $\Qe{i}{j}{}{}$ is a $|\T|$-by-$|\T|$ matrix that gives the probabilities of node $j$ to transmit a packet sent by node $i$ for all possible combinations of time slot. Its structure is:
\begin{equation}
\Qe{i}{j}{}{} = \left[
\begin{array}{ccc}
\Qe{i}{j}{1}{1}	&	\cdots	& \Qe{i}{j}{1}{|\T|} \\
\vdots		& 			&	 \vdots \\
\Qe{i}{j}{|\T|}{1}& 	\cdots	&	 \Qe{i}{j}{|\T|}{|\T|} \\
\end{array}
\right]
\label{eq:relaying2}
\end{equation}
\noindent where $\Qe{i}{j}{u}{v}$ is the probability for a node $j$ to retransmit on channel $v$ a packet that has been transmitted by node $i$ on time slot $u$. With respect to our network model, it is equal to:
\[\Qe{i}{j}{u}{v} = p_{ij}^{u} (1-\tau_j^v) x_{ij}^{uv}\] 
It is important to note that the powers of matrix $\Qmatrix$ tend to 0 if its elements verify the strict inequality $\Qe{i}{j}{u}{v} <1$. For this property to hold, we have to ensure that $p_{ij}^{u}<1$. This is usually the case for a wireless transmission where perfect channels exist in very few unrealistic cases.

\emph{The arrival matrix $\Rmatrix$} is given by:
\[
\Rmatrix = \left[
\begin{array}{ccc}
\Qd{1}{D_1}{}{}	& \cdots 	& 	\Qd{1}{D_{|\D|}}{}{} \\
\vdots		&		&	\vdots	\\
\Qd{N}{D_1}{}{}	& \cdots 	& 	\Qd{N}{D_{|\D|}}{}{} \\
\end{array}
\right]
\]
\noindent $\Qd{i}{D_j}{}{}$ is a $|\T|$-by-$|\T|$ diagonal matrix whose diagonal elements $\Qd{i}{D_j}{u}{}$ give the probabilities for a packet transmitted by a node $i$ in time slot $u$ to arrive at destination $D_j$.  Consequently, $\Qd{i}{D_j}{u}{}$ is the probability of reception:
\[\Qd{i}{D_j}{u}{}=p_{iD_j}^{u} (1-\tau_{D_j}^v) =p_{iD_j}^{u} \]
\noindent since $\tau_{D_j}^v=0, \forall v \in \T$ for any destination node $D_j$.

\emph{The source matrix $\Smatrix$} is needed to compute the initial flow vector $\Fout{out}{}{1}$, naming the probabilities of all the nodes to transmit a packet sent by a source. $\Smatrix$ is a $|\Orig|$-by-$|\T|$ matrix. An element $\tau_{S_{i}}^u$ gives the probability of the source $i$ to transmit in time slot $u$. An initial flow matrix is obtained from $\Smatrix$ with

\[
\Fout{src}{}{1} = \Smatrix \Msrc
\]       
\noindent where $\Msrc$ is a $|\T|$-by-$(N+|\D|).|\T|$ transition matrix defined as $\Msrc=\left[ \Qsrc ~ \Rsrc \right]$ with $\Qsrc$ and $\Rsrc$ the relaying and arrival matrix for the packets sent by the source, respectively. We have 
\[
\Qsrc =  \left[
\begin{array}{ccc}
\Qe{S}{1}{}{}	& \cdots 	& 	\Qe{S}{N}{}{} \\
\end{array}
\right]
\]       
\noindent and 
\[
\Rsrc =  \left[
\begin{array}{ccc}
\Qd{S}{D_1}{}{}	& \cdots 	& 	\Qd{S}{D_{|\D|}}{}{} \\
\end{array}
\right]
\]       
\noindent where $\Qe{S}{i}{}{}$ follows the pattern given by \eqref{eq:relaying2}. Since destinations do not retransmit packets, $D_{SD_i}$ is a $|\T|$-by-$|\T|$ diagonal matrix whose diagonal elements are $D_{SD_i}^u = p_{SD_i}^u, u \in \T$.

The flow matrix $\Fout{src}{}{1}$ created at epoch $s=1$ is a $|\Orig|$-by-$(N+|\D|).|\T|$ matrix of flows. The initial flow vector $\Fout{out}{}{1}$ relative to source $i$ is given by the $i^{th}$ line of $\Fout{src}{}{1}$. 
A very interesting feature of the steady state performance evaluation and network model proposed herein is that the matrix $\Qmatrix$ and $\Rmatrix$ can be applied to any type of source-destination communication (unicast, multicast, anycast, ...). Indeed, this is a direct consequence of the fact that the inherent broadcast and interference-limited properties of the radio channel are accounted for in the structure of $\Qmatrix$.  
For instance, for a multicast transmission, the source matrix $\Smatrix$ is composed of a single source and the arrival matrix  $\Rmatrix$ is defined for  $|\D|>1$ destinations. The matrix $\Qmatrix$ stays unchanged.
Or when multiple sources are transmitting to a same destination (a sink for instance in a wireless sensor network), a source matrix of $\Smatrix$ is composed of several nodes. Each one of them is propagated in the network using the same $\Qmatrix$ to a unique destination modeled by $\Rmatrix$.  

For clarity purposes, we have presented all matrices using all $N$ possible relays. However, when considering a specific solution $\tau$, these matrices are derived only for the set $\M$ of active transmissions. For instance, the size of the square matrix $\Qmatrix$ is reduced from $l=N.|\T|$ to $l=\M$.  

\subsection{Fundamental matrix}

We really want to stress that the proposed transition matrix definition is not the definition of a Markov chain. Indeed, the sum of the probabilities on a line of $M$ is higher than $1$. However, having $\Qe{i}{j}{u}{v}<1$ ensures that $\Qmatrix^s$ tends to zero as the number of time epochs $s$ tends to infinity. 
As a consequence, the following theorem still holds:

{\sc Theorem 1:} $I-\Qmatrix$ has an inverse, and
\[
(I-\Qmatrix)^{-1} = I + Q + Q^2+ \cdots = \sum_{s=0}^{\infty} Q^s
\]

{\sc Proof}
The complete proof is omitted for conciseness purposes but can be found in \cite{markov}. Even though $\Mmatrix$ is not a Markovian transition matrix, the only property needed for Theorem 1 to hold in our case is that $\Qmatrix^k \rightarrow 0$ as $k \rightarrow \infty$. 

\hspace{4.2in}{\textit{q.e.d.}

The inverse of $I-\Qmatrix$ is defined as the \emph{fundamental matrix} and denoted $\Nmatrix = (I-\Qmatrix)^{-1}$.
In our case, the fundamental matrix gives us the mean of the total number of transmissions that have been done until all packets be received at the destination. 

\subsection{Optimization criteria}

The definition of our optimization criteria are directly derived from the fundamental matrix. Indeed, if $I-\Qmatrix$ is invertible, $\Nmatrix$ gives a measure of the performance of one solution $X \in \X$ when the number of time epochs tends to infinity. The very interesting feature is that it also accounts for all the possible cycles in the graph and consequently models the broadcast property of the wireless channel.  
The criteria here after are defined for one source-destination flow. They will be extended in future work to multiple flows which is possible since the complete framework can model concurrent flows on the network.  

\paragraph{Redundancy, capacity and robustness} It is possible to derive for one flow the number of packets received at destination for one transmitted packet  ($|\Smatrix|=1$) with 
\[
f = \Fout{out}{}{1} \cdot 
	\left[ \begin{array}{cc} 
		\Nmatrix & \Rmatrix \\ 
		0	& 	I \\ 
	\end{array} \right] 
\cdot  \left[ 
	\begin{array}{c} 
		0  \\ 
		1	 \\ 
	\end{array}
\right]
\]  
\noindent where $ \left[ 
	\begin{array}{cc} 
		0^\dagger  &
		1^\dagger	 \\ 
	\end{array}
\right]^\dagger$ is a vector of $N|\T|$ zeros followed by $|\D||\T|$ ones that accumulate all packets being received at the destinations into $f$. 
This model shows that a solution $X \in \X$ provides $f$ packets at the destination for one sent packet. However, since relays are not able to discriminate packets, the set of $f$ packets may be composed in the worst case of $f$ copies of a packet by the source. What we can say here is that:

$\bullet$ if $f\geq1$, there may be at most $f$ different packets. Consequently, $f$ provides both an upper bound on the network capacity and redundancy. Let $f_C$ be the bound on the real capacity of the network. It is given by
\[ f_C = \min(1, f)
\]  This bound can be reached if all the packets received at the source are orthogonal. This may be possible if an optimal network coding strategy is found where all transmitted packets are encoded combinations of the original information. 

$\bullet$ if $f<1$, $f$ can be interpreted as a robustness criterion giving the probability to receive one message at destination.      

This criterion can be used to remove solutions that do not guaranty robustness. For instance, solutions having $f<1$ can be disregarded if robustness wants to be guaranteed.

\paragraph{Delay criterion}
Let $p(H=h)$ be the probability for a transmission to be done in $h$ hops. Thus 
\[
p(H=h) = \left\{ 
	\begin{array}{cc}
		\Fout{out}{}{1} \cdot \left[ \begin{array}{c} \Rmatrix \\ I \\ \end{array} \right]  & h=2 \\
		\Fout{out}{}{1} \cdot \left[ \begin{array}{c} \Qmatrix \\ 0 \\ \end{array} \right]  \cdot \left[ \begin{array}{c} \Rmatrix \\ I \\ \end{array} \right]  & h>2 \\		
	\end{array}
\right. 
\]
We assume that a relay introduces a delay of 1 unit. Consequently, a $h$-hop transmission introduces a delay of $h-1$ units.  
It is given by the infinite sum: 
\[
\begin{split}
f_D = \sum_{h=1}^{\infty} (h-1)\cdot p(H=h) = \\
  \Fout{out}{}{1} \cdot 
  \left[ 
	\begin{array}{c}  
  	Id + 2\Qmatrix +  \cdot\cdot + (h+1)\Qmatrix^h + \cdot\cdot \\
  	0 \\
	\end{array}
  \right]	
  \cdot\left[ \begin{array}{c} \Rmatrix \\ I \\ \end{array} \right] \cdot 
 \left[ 
	\begin{array}{c} 
		0  \\ 
		1	 \\ 
	\end{array}
\right]
\end{split}
\]
Let $V$ be equal to $(Id + 2\Qmatrix +\cdot\cdot + (h+1)\Qmatrix^h + \cdot\cdot) $. We can show that $V=(\Nmatrix)^2$ and consequently
\[
f_D = \Fout{out}{}{1} \cdot 
	\left[ \begin{array}{cc} 
		(\Nmatrix)^2 & \Rmatrix \\ 
		0	& 	I \\ 
	\end{array} \right] 
\cdot  \left[ 
	\begin{array}{c} 
		0  \\ 
		1	 \\ 
	\end{array}
\right]
\] 
{\sc Proof}
We can substitute $Id$ by $\Qmatrix + \Nmatrix$ in the definition of $V$. Then we get by factorizing $\Qmatrix$:
\[
V = \Nmatrix + \Qmatrix \cdot(Id + 2\Qmatrix +  3\Qmatrix^2 + \cdots + (h+1)\Qmatrix^h + \cdots ) 
\]
We now have $V=\Nmatrix + \Qmatrix \cdot V$. Consequently $V=(\Nmatrix)^2$.

\hspace{4.2in}{\it q.e.d.}

\paragraph{Energy criterion}

The energy is measured by the number of packets being transmitted in the network. It can be simply derived by:

\[
f_E = \Fout{out}{}{1} \cdot 
	\left[ \begin{array}{c} 
		\Nmatrix  \\ 
		0	 \\ 
	\end{array} \right] 
\cdot \left[ 
	\begin{array}{c} 
		1  \\ 
		0	 \\ 
	\end{array}
\right]
\]
\noindent where $1$ is a $N|\T|$-by-$1$ vector of ones that sums all the packets sent by the relays that have participated in the transmission.

\subsection{Example: 1-relay network}

In this example, we consider a basic 1-relay network as the one depicted in Fig.~\ref{fig:1relay}.
\begin{figure}
\begin{center}
  \scalebox{0.5}{
  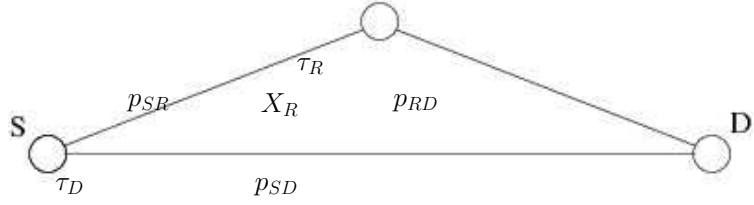}
   \caption{1-Relay network}
\label{fig:1relay}
\end{center}
\end{figure}
Interference is completely mitigated in this network if we use $|\T|=2$ times slots and $S$ transmits in time slot 1 and relay $R$ in time slot 2. As a consequence we have $\Smatrix=\left[\tau_S^1~~ 0 \right]$ and $\tau_R = \left[0 ~~ \tau_R^2 \right]$. The forwarding probabilities matrix is:
\[X_R = \left[ 
\begin{array}{cc}
	x_{SR}^{11} & x_{SR}^{12} \\ 
	x_{SR}^{21} & x_{SR}^{22} \\
\end{array}
\right] = 
\left[ 
\begin{array}{cc}
	0 & \tau_R^2 / (\tau_S^1.p_{SR}^1.(1-\tau_R^2)) \\ 
	0 & 0 \\
\end{array}
\right] 
\]  
where the values of the $x_{ij}^{uv}$ are derived according to \eqref{eq:fwfprobaflowconservation}. 
The derivation of $\Fout{out}{}{1} = \Smatrix \Msrc$ provides:
\[
 \Fout{out}{}{1}  = \left[ 
\begin{array}{ccc}
	\tau_S^1.p_{SR}^1.(1-\tau_R^2).x_{SR}^{12} &
	\tau_S^1.p_{SD}^1 &
	0\\
\end{array}
\right]
\]
Since we have only one relay, matrix $\Qmatrix$ does not exist and we have $\Mmatrix = \left[ D^\dagger~~I^\dagger \right]^\dagger$. The criteria $f$ equals:
\[
f = \Fout{out}{}{1}\cdot \left[ \begin{array}{c} D\\ I \end{array} \right] 
\cdot \left[ 
	\begin{array}{c} 
		1  \\ 
		1	 \\ 
	\end{array}
\right]
= \Fout{out}{}{1} \cdot \left[ 
\begin{array}{ccc}
	0	& 	p_{RD}^2 \\
	1	&	0 \\
	0	&	1 \\
\end{array}
\right] \cdot \left[ 
	\begin{array}{c} 
		1  \\ 
		1	 \\ 
	\end{array}
\right]
\]
\noindent where $f$ is obtained for $\tau_S^1=1$. The bound on capacity of this 1-relay network can be obtained for $f=1$. 
Thus, we can derive the value of $(\tau_R^2)^*$ that guaranties $f=1$ and verifies: 
\[
p_{SD}^1+(\tau_R^2)^*.p_{RD}^2 =1
\]
\noindent It shows that the optimal transmission rate of the relay with respect to capacity is the rate at which the transmissions on the link $(R,D)$ compensate the losses on the link $(S,D)$. 
However, this bound may not be achievable when the sum of the information coming into $D$ and $R$ is lower than 1. 
Typically, applying a cut-set between $S$ and the set $\{R,D\}$, leads to a maximal flow $f \leq 1-(1-p_{SD}^1)\cdot(1-p_{SR}^1)$.
Further, this capacity bound can only be achieved if the relay knows which packet have not been received by $D$ directly from $S$ or, more interestingly, by performing a network coding to introduce proper diversity in the flow.

\section{Conclusion and discussion}\label{sec:conclusion}
The work presented in this paper has derived a flexible and powerful framework for evaluating the performance of a wireless ad hoc network with respect to several performance criteria. It has been designed to account for the broadcast nature of wireless communications and for an accurate interference temperature characterization for the network. The proposed framework allows for the determination of Pareto-optimal solutions with respect to capacity, redundancy, delay and energy optimization criteria. 
The solutions defined herein represent a steady state of the network where all active relays are constantly transmitting data. 
  
The complexity of the proposed framework is not a function of the number or the structure of the source-destination flows and it is scalable with respect to the number of flows in the network. As a consequence, it can model various types of transmission: one-to-one, one-to-many, many-to-one or many-to-many.

This flexibility comes at the price of an extended problem search space. However, the structure of this search space is very interesting from the optimization point of view since there is a dependency between the set of feasible transmission rate matrices $\Gamma$ and the set of feasible forwarding probabilities matrices $\X_f$. If it is possible to disregard a feasible value of $\tau$ because it is not promising enough for the MO search, then the worst case complexity of the problem can be reduced from $O(|\Delta|^{N^2.|\T|})$ to  $O(|\Delta|^{N.|\T|})$.

The definition of the fundamental matrix permits to derive an upper bound on the capacity of the network. This upper bound is provided in terms of the average number of packets being received at the destination. With our framework, we can not differentiate packets at the destination and hence can not state whether a received packets is a copy of an already received packet of not. However, we state that using proper network coding techniques, it is possible to get an network effective capacity close enough from this bound.  

This paper has focused on presenting theoretically our MO performance evaluation framework. It is only illustrated by a basic example. In future work, we will have to derive proper MO optimization search algorithms \cite{siarry} to efficiently compute the set of Pareto-optimal solutions. This task is challenging but as mentioned in the paper, it is often enough to address the $N_{max}$-relaying subproblem (with $N_{max}<N$ the maximum number of active relays in the network). Next, it will be of major interest to analyze the Pareto-optimal bounds and its corresponding solutions. The knowledge of such bounds will provide a powerful tool to assess the quality of distributed routing and resource allocation algorithms. Further, and this is of great interest in our opinion, we can take advantage of the steady state Pareto-optimal solutions to design distributed protocols that get close to our MO performance bound.

\section*{Acknowledgments}
This work was supported in part by the Marie Curie IOF Action of the European Community's Sixth Framework Program (DistMO4WNet project). This article only reflects the authors' views and the European Community are liable for any use that may be made of the information contained herein.

\bibliographystyle{IEEEtran}
\bibliography{IEEEabrv,basebiblio}

\tableofcontents

\end{document}